\documentclass[12pt,reqno]{amsart}
\usepackage{graphicx}
\usepackage{amscd}
\usepackage{amsmath}
\usepackage{epsfig}
\usepackage{amsfonts}
\usepackage{amssymb}

\providecommand{\U}[1]{\protect\rule{.1in}{.1in}}
\providecommand{\U}[1]{\protect\rule{.1in}{.1in}}
\allowdisplaybreaks

\theoremstyle{plain}

\numberwithin{equation}{section}

\input{tcilatex}

\begin{document}
\title[Relativistic Kramers--Pasternack Recurrence Relations]{Relativistic
Kramers--Pasternack Recurrence Relations}
\author{Sergei K. Suslov}
\address{School of Mathematical and Statistical Sciences and Mathematical,
Computational and Modeling Sciences Center, Arizona State University, Tempe,
AZ 85287--1804, U.S.A.}
\email{sks@asu.edu}
\urladdr{http://hahn.la.asu.edu/\symbol{126}suslov/index.html}
\date{\today }
\subjclass{Primary 81Q05. Secondary 33C20}
\keywords{The Dirac equation, hydrogenlike ions, expectation values, Hahn
polynomials, Kramers--Pasternack recurrence relation}

\begin{abstract}
Recently we have evaluated the matrix elements $\langle Or^{p}\rangle ,$
where $O$ $=\left\{ 1,\beta ,i\mathbf{\alpha n}\beta \right\} $ are the
standard Dirac matrix operators and the angular brackets denote the
quantum-mechanical average for the relativistic Coulomb problem, in terms of
generalized hypergeometric functions $\ _{3}F_{2}\left( 1\right) $ for all
suitable powers and established two sets of Pasternack-type matrix
identities for these integrals. The corresponding Kramers--Pasternack type
three-term vector recurrence relations are derived here.
\end{abstract}

\maketitle

\section{Introduction}

Recent experimental and theoretical progress has renewed interest in quantum
electrodynamics of atomic hydrogenlike systems (see, for example, \cite%
{Gum05}, \cite{Gum07}, \cite{Karsh01}, \cite{Kash03}, \cite{Mohr:Plun:Soff98}%
, \cite{ShabGreen}, and \cite{ShabYFN08} and references therein). In the
last decade, the two-time Green's function method of deriving formal
expressions for the energy shift of a bound-state level of high-$Z$
few-electron systems was developed \cite{ShabGreen} and numerical
calculations of QED effects in heavy ions were performed with excellent
agreement to current experimental data \cite{Gum05}, \cite{Gum07}, \cite%
{ShabYFN08}. These advances motivate, among other technical things,
investigation of the expectation values of the Dirac matrix operators
between the bound-state relativistic Coulomb wave functions. Special cases
appear in calculations of the magnetic dipole hyperfine splitting, the
electric quadrupole hyperfine splitting, the anomalous Zeeman effect, and
the relativistic recoil corrections in hydrogenlike ions (see, for example, 
\cite{Adkins}, \cite{ShabHyd}, \cite{ShabHydVir}, \cite{Suslov} and
references therein).\medskip

In the previous paper \cite{Suslov}, the matrix elements $\langle
Or^{p}\rangle ,$ where $O$ $=\left\{ 1,\beta ,i\mathbf{\alpha n}\beta
\right\} $ are the standard Dirac matrix operators and the angular brackets
denote the quantum-mechanical average for the relativistic Coulomb problem,
have been evaluated as sums of three special generalized hypergeometric
functions $\ _{3}F_{2}\left( 1\right) $ (or Chebyshev polynomials of a
discrete variable) for all suitable powers. As a result, two sets of
Pasternack-type matrix symmetry relations for these integrals, when $%
p\rightarrow -p-1$ and $p\rightarrow -p-3,$ have been derived. We
concentrate on what are essentially radial integrals since, for problems
involving spherical symmetry, one can reduce all expectation values to
radial integrals by use of the properties of angular momentum.\medskip

Our next goal is to establish relativistic analogs of the
Kramers--Pasternack recurrence relation \cite{Kram}, \cite{Past}, and \cite%
{PastCH} (some progress in this direction had been made in \cite{Andrae97}).
Here several three-term vector recurrence relations are obtained, which
follow in a natural way from the well-known theory of classical orthogonal
polynomials of a discrete variable \cite{Ni:Su:Uv} and basic matrix algebra.
This paper is organized as follows. We review the nonrelativistic case in
the next section and then present relativistic extensions, that seem to be
new, in sections~3 and 4. The last section contains a new interpretation of
the known two-term recurrence relations for these relativistic expectation
values \cite{ShabVest}, \cite{Shab91}. The new recurrence relations are
summarized in appendix~A in matrix form for the benefit of the
reader.\medskip 

The author believes that these mathematical results are not only natural and
elegant but also will be useful in the current theory of hydrogenlike heavy
ions and other exotic relativistic Coulomb systems. At the same time, the
mathematical structure behind these expectation values remains not fully
understood \cite{Andrae97}. For example, to the best of my knowledge,
numerous recurrence relations for the radial integrals, obtained with the
help of a hypervirial theorem (see \cite{Adkins}, \cite{ShabVest}, \cite%
{Shab91}, \cite{ShabHydVir} and references therein), do not follow in an
obvious way from the advanced theory of generalized hypergeometric functions 
\cite{Ba}. After more than 80 years of a thorough investigation, the
relativistic Coulomb problem keeps generating some mathematical challenges.

\section{Expectation Values $\langle r^{p}\rangle $ for Nonrelativistic
Coulomb Problem}

Evaluation of the matrix elements $\langle r^{p}\rangle $ between
nonrelativistic bound-state hydrogenlike wave functions has a long history
in quantum mechanics. An incomplete list of references includes \cite%
{Andrae97}, \cite{Armstr71a}, \cite{Balas00}, \cite{Ba:Zel:Per}, \cite{Beker}%
, \cite{Be:Sal}, \cite{Blanch74}, \cite{Bock74}, \cite{Chac:Lev:Mosh76}, 
\cite{Curt91}, \cite{Drake:Swain90}, \cite{Ep:Ep}, \cite{HeyAug93}, \cite%
{Kram}, \cite{La:Lif}, \cite{Libo}, \cite{Marx91}, \cite{Mes}, \cite{More83}%
, \cite{Ojha:Croth84}, \cite{Past}, \cite{PastCH}, \cite{Per:Zel}, \cite%
{Qiang:Dong04}, \cite{Shab91}, \cite{Shertzer}, \cite{Sus:Trey}, \cite%
{Swain:Drake90}, \cite{VanVleck34}, and \cite{Waller}. Different methods
were used in order to derive these matrix elements. For example, in Ref.~%
\cite{Sus:Trey} (see also \cite{Past}, \cite{PastCH}) the mean values for
states of definite energy have been obtained in terms of the Chebyshev
polynomials of a discrete variable $t_{m}\left( x,N\right) =h_{m}^{\left(
0,\ 0\right) }\left( x,N\right) $ originally introduced in Refs.~\cite%
{Chebyshev59} and \cite{Chebyshev64}. The so-called Hahn polynomials,
introduced also by P.~L.~Chebyshev \cite{Chebyshev75} and given by%
\begin{equation}
h_{m}^{\left( \alpha ,\ \beta \right) }\left( x,N\right) =\frac{\left(
1-N\right) _{m}\left( \beta +1\right) _{m}}{m!\;}\ _{3}F_{2}\left( 
\begin{array}{c}
-m\medskip ,\ \alpha +\beta +m+1,\ -x \\ 
\beta +1\medskip ,\quad 1-N%
\end{array}%
;\ 1\right) ,  \label{hahn}
\end{equation}%
were rediscovered and generalized in the late 1940s by W.~Hahn (we use the
standard definition of the generalized hypergeometric series throughout the
paper \cite{Ba}, \cite{Erd}).\medskip

The final results have the following closed forms%
\begin{equation}
\left\langle r^{k-1}\right\rangle =\frac{1}{2n}\left( \frac{na_{0}}{2Z}%
\right) ^{k-1}t_{k}\left( n-l-1,-2l-1\right) ,  \label{in2}
\end{equation}%
when $k=0,1,2,...$ and%
\begin{equation}
\left\langle \frac{1}{r^{k+2}}\right\rangle =\frac{1}{2n}\left( \frac{2Z}{%
na_{0}}\right) ^{k+2}t_{k}\left( n-l-1,-2l-1\right) ,  \label{in3}
\end{equation}%
when $k=0,1,...,\;2l.$ Here $a_{0}=\hslash ^{2}/me^{2}$ is the Bohr radius
(more details can be found in Refs.~\cite{Cor-Sot:Sus} and \cite{Sus:Trey}%
).\medskip

The ease of handling of these matrix elements for the discrete levels is
greatly increased if use is made of the known properties of these classical
polynomials of Chebyshev \cite{Erd}, \cite{Karlin:McGregor61}, \cite{Ko:Sw}, 
\cite{Ni:Su:Uv}, \cite{Ni:Uv} and \cite{Chebyshev59}, \cite{Chebyshev64}, 
\cite{Chebyshev75}. The direct consequences of (\ref{in2})--(\ref{in3}) are
an inversion\ relation:%
\begin{equation}
\left\langle \frac{1}{r^{k+2}}\right\rangle =\left( \frac{2Z}{na_{0}}\right)
^{2k+1}\frac{\left( 2l-k\right) !}{\left( 2l+k+1\right) !}\ \left\langle
r^{k-1}\right\rangle  \label{in4}
\end{equation}%
with $0\leq k\leq 2l$ and the three-term recurrence relation:%
\begin{equation}
\left\langle r^{k}\right\rangle =\frac{2n\left( 2k+1\right) }{k+1}\left( 
\frac{na_{0}}{2Z}\right) \left\langle r^{k-1}\right\rangle -\frac{k\left(
\left( 2l+1\right) ^{2}-k^{2}\right) }{k+1}\left( \frac{na_{0}}{2Z}\right)
^{2}\left\langle r^{k-2}\right\rangle  \label{in5}
\end{equation}%
with the initial conditions%
\begin{equation}
\left\langle \frac{1}{r}\right\rangle =\frac{Z}{a_{0}n^{2}},\qquad
\left\langle 1\right\rangle =1,  \label{in6}
\end{equation}%
which is convenient for evaluation of the mean values $\left\langle
r^{k}\right\rangle $ for $k\geq 1.$\medskip

In our approach, the recurrence relation (\ref{in5}) is a special case of
the three-term recurrence relation for the Hahn polynomials (\ref{3term}).
It was originally found by Kramers and Pasternack in the late 1930s \cite%
{Kram}, \cite{Past}, and \cite{PastCH}. The inversion symmetry (\ref{in4}),
which is also due to Pasternack, has been rediscovered many years later \cite%
{Bock74}, \cite{More83} (see also \cite{HeyJan93} and \cite{HeyAug93} for
historical comments). Generalizations of (\ref{in4})--(\ref{in5}) for
off-diagonal matrix elements are discussed in Refs.~\cite{Blanch74}, \cite%
{Ding87}, \cite{Ep:Ep:Ken}, \cite{HeyAug93}, \cite{Mor:Pin:Tip91}, \cite%
{Nu:Lop:Sal95}, \cite{Ojha:Croth84}, \cite{Past:Stern}, \cite{San:Mor:Pin92}%
, \cite{ShabVest}, \cite{Shab91}, \cite{Shertzer}, and \cite{Swain:Drake90}.
The properties of the hydrogenlike \ radial matrix elements are considered
from a group-theoretical viewpoint in Refs.~\cite{Armstr71a}, \cite%
{Chac:Lev:Mosh76}, and \cite{Ojha:Croth84}. Extensions to the relativistic
case are given in \cite{Adkins}, \cite{Andrae97}, \cite{Davis}, \cite{Ep:Ep}%
, \cite{Owono:Kwato:Oumarou}, \cite{Suslov}, and \cite{Sus:Trey} (see also
references therein and the following sections of this paper).\medskip 

In a retrospect, Pasternack's papers \cite{Past}, \cite{PastCH} had paved
the road to the discovery of the continuous Hahn polynomials in the mid
1980s (see \cite{AskeyCH}, \cite{At:SusCH}, \cite{KoelinkCH}, \cite{Sus:Trey}
and references therein).\medskip

\section{Three-Term Recurrence Relations for Relativistic Matrix Elements}

In the present paper we establish relativistic analogs of the
Kramers--Pasternack recurrence relations (\ref{in5}) for the following set
of integrals of the relativistic radial functions:%
\begin{eqnarray}
A_{p} &=&\int_{0}^{\infty }r^{p+2}\left( F^{2}\left( r\right) +G^{2}\left(
r\right) \right) \ dr,  \label{meA} \\
B_{p} &=&\int_{0}^{\infty }r^{p+2}\left( F^{2}\left( r\right) -G^{2}\left(
r\right) \right) \ dr,  \label{meB} \\
C_{p} &=&\int_{0}^{\infty }r^{p+2}F\left( r\right) G\left( r\right) \ dr.
\label{meC}
\end{eqnarray}%
With the notations from Refs.~\cite{Suslov} and \cite{Sus:Trey}, one can
evaluate these integrals in terms of the Chebyshev and Hahn polynomials of a
discrete variable and present the answer, say, when $p\geq 0,$ in the
following closed form:%
\begin{eqnarray}
&&4\mu \nu ^{2}\left( 2a\beta \right) ^{p}\ A_{p}  \label{meAf} \\
&&\quad =a\kappa \left( \varepsilon \kappa +\nu \right) \ h_{p+1}^{\left(
0,\ 0\right) }\left( n-1,-1-2\nu \right)  \notag \\
&&\quad -2\frac{p+2}{p+1}\mu \left( \varepsilon ^{2}\kappa ^{2}-\nu
^{2}\right) \ h_{p}^{\left( 1,\ 1\right) }\left( n-1,-1-2\nu \right)  \notag
\\
&&\quad +a\kappa \left( \varepsilon \kappa -\nu \right) \ h_{p+1}^{\left(
0,\ 0\right) }\left( n,1-2\nu \right) ,  \notag
\end{eqnarray}%
\begin{eqnarray}
&&4\mu \nu \left( 2a\beta \right) ^{p}\ B_{p}  \label{meBf} \\
&&\quad =a\left( \varepsilon \kappa +\nu \right) \ h_{p+1}^{\left( 0,\
0\right) }\left( n-1,-1-2\nu \right)  \notag \\
&&\quad -a\left( \varepsilon \kappa -\nu \right) \ h_{p+1}^{\left( 0,\
0\right) }\left( n,1-2\nu \right) ,  \notag
\end{eqnarray}%
\begin{eqnarray}
&&8\mu \nu ^{2}\left( 2a\beta \right) ^{p}\ C_{p}  \label{meCf} \\
&&\quad =a\mu \left( \varepsilon \kappa +\nu \right) \ h_{p+1}^{\left( 0,\
0\right) }\left( n-1,-1-2\nu \right)  \notag \\
&&\quad -2\frac{p+2}{p+1}\kappa \left( \varepsilon ^{2}\kappa ^{2}-\nu
^{2}\right) \ h_{p}^{\left( 1,\ 1\right) }\left( n-1,-1-2\nu \right)  \notag
\\
&&\quad +a\mu \left( \varepsilon \kappa -\nu \right) \ h_{p+1}^{\left( 0,\
0\right) }\left( n,1-2\nu \right) .  \notag
\end{eqnarray}%
Here,%
\begin{eqnarray}
&&\kappa =\pm \left( j+1/2\right) ,\qquad \nu =\sqrt{\kappa ^{2}-\mu ^{2}}, 
\notag \\
&&\mu =\alpha Z=Ze^{2}/\hbar c,\qquad a=\sqrt{1-\varepsilon ^{2}},
\label{notations} \\
&&\varepsilon =E/mc^{2},\qquad \beta =mc/\hbar  \notag
\end{eqnarray}%
with the total angular momentum $j=1/2,3/2,5/2,\ ...$ (see \cite{Suslov} and 
\cite{Sus:Trey} for more details).\medskip

Although a set of useful recurrence relations between the relativistic
matrix elements was derived by Shabaev \cite{ShabVest}, \cite{Shab91} (see
also \cite{Adkins}, \cite{Ep:Ep}, \cite{ShabHydVir}, \cite{Suslov}, \cite%
{Vrs:Ham}, and the last section of this paper) on the basis of a hypervirial
theorem, the corresponding relativistic Kramers--Pasternack type relations
seem to be missing in the available literature. Our equations (\ref{meAf})--(%
\ref{meCf}) reveal that they should have a vector form.\medskip\ 

Here, one can apply a familiar three-term recurrence relation for the Hahn
polynomials \cite{Ni:Su:Uv}, \cite{Ni:Uv},%
\begin{equation}
xh_{m}^{\left( \alpha ,\ \beta \right) }\left( x,N\right) =\alpha
_{m}h_{m+1}^{\left( \alpha ,\ \beta \right) }\left( x,N\right) +\beta
_{m}h_{m}^{\left( \alpha ,\ \beta \right) }\left( x,N\right) +\gamma
_{m}h_{m-1}^{\left( \alpha ,\ \beta \right) }\left( x,N\right)  \label{3term}
\end{equation}%
with%
\begin{eqnarray}
\alpha _{m} &=&\frac{\left( m+1\right) \left( \alpha +\beta +m+1\right) }{%
\left( \alpha +\beta +2m+1\right) \left( \alpha +\beta +2m+2\right) },
\label{3terma} \\
\beta _{m} &=&\frac{\alpha -\beta +2N-2}{4}+\frac{\left( \beta ^{2}-\alpha
^{2}\right) \left( \alpha +\beta +2N\right) }{4\left( \alpha +\beta
+2m\right) \left( \alpha +\beta +2m+2\right) },  \label{3termb} \\
\gamma _{m} &=&\frac{\left( \alpha +m\right) \left( \beta +m\right) \left(
\alpha +\beta +N+m\right) \left( N-m\right) }{\left( \alpha +\beta
+2m\right) \left( \alpha +\beta +2m+1\right) }  \label{3termc}
\end{eqnarray}%
three times for the corresponding special cases in (\ref{meAf})--(\ref{meCf}%
). Introducing two sets of vectors%
\begin{equation}
\mathbf{A}_{p}=\left( 
\begin{array}{c}
A_{p}\smallskip \\ 
B_{p}\smallskip \\ 
C_{p}%
\end{array}%
\right) ,\qquad \mathbf{X}_{p}=\left( 
\begin{array}{c}
\ h_{p+1}^{\left( 0,\ 0\right) }\left( n-1,-1-2\nu \right) \medskip \\ 
\frac{p+2}{p+1}h_{p}^{\left( 1,\ 1\right) }\left( n-1,-1-2\nu \right)
\smallskip \medskip \\ 
h_{p+1}^{\left( 0,\ 0\right) }\left( n,1-2\nu \right)%
\end{array}%
\right) ,  \label{vectors}
\end{equation}%
we conclude that the recurrence relations in question should have the
following matrix structure:%
\begin{equation}
\mathbf{A}_{p}=T\mathbf{X}_{p},\qquad \mathbf{X}_{p}=T^{-1}\mathbf{A}_{p}
\label{mat}
\end{equation}%
with%
\begin{equation}
\mathbf{X}_{p+1}=U_{p}\mathbf{X}_{p}+V_{p}\mathbf{X}_{p-1},  \label{matrec}
\end{equation}%
where $U_{p}$ and $W_{p}$ are known diagonal matrices. Then%
\begin{equation}
\mathbf{A}_{p+1}=D_{p}\mathbf{A}_{p}+E_{p}\mathbf{A}_{p-1},  \label{matreca}
\end{equation}%
where%
\begin{equation}
D_{p}=TU_{p}T^{-1},\qquad E_{p}=TV_{p}T^{-1}  \label{simmat}
\end{equation}%
are similar matrices.\medskip

The elementary linear algebra argument outlined above allows us to obtain a
relativistic generalization of the Kramers--Pasternack recurrence relation (%
\ref{in5}) as follows: 
\begin{eqnarray}
\left( 2a\beta \right) ^{2}\ A_{p+1} &=&4\beta \varepsilon \mu \frac{2p+3}{%
p+2}\ A_{p}  \label{4termA} \\
&&-\frac{p\left( p+2\right) \left( 4\nu ^{2}-\left( p+1\right) ^{2}\right)
+4\kappa ^{2}}{\left( p+1\right) \left( p+2\right) }\ A_{p-1}  \notag \\
&&-4\kappa \frac{p+1}{p+2}\ B_{p-1}+\frac{8\kappa \mu }{\left( p+1\right)
\left( p+2\right) }\ C_{p-1},  \notag
\end{eqnarray}%
\begin{eqnarray}
\left( 2a\beta \right) ^{2}\ B_{p+1} &=&4\beta \varepsilon \mu \frac{2p+3}{%
p+2}\ B_{p}  \label{4termB} \\
&&-\left( 4\nu ^{2}-p\left( p+2\right) \right) \frac{p+1}{p+2}\ B_{p-1} 
\notag \\
&&-4\kappa \frac{p+1}{p+2}\ A_{p-1}+8\mu \frac{p+1}{p+2}\ C_{p-1},  \notag
\end{eqnarray}%
\begin{eqnarray}
\left( 2a\beta \right) ^{2}\ C_{p+1} &=&4\beta \varepsilon \mu \frac{2p+3}{%
p+2}\ C_{p}  \label{4termC} \\
&&-\frac{2\kappa \mu }{\left( p+1\right) \left( p+2\right) }\ A_{p-1}-2\mu 
\frac{p+1}{p+2}\ B_{p-1}  \notag \\
&&-\frac{p\left( p+2\right) \left( 4\nu ^{2}-\left( p+1\right) ^{2}\right)
-4\mu ^{2}}{\left( p+1\right) \left( p+2\right) }\ C_{p-1}.  \notag
\end{eqnarray}%
Our relations (\ref{4termA})--(\ref{4termC}), together with the symmetry
under the reflections, $p\rightarrow -p-1$ and $p\rightarrow -p-3,$
established in \cite{Suslov}, give a possibility to compute all the required
matrix elements in a recursive manner without direct evaluation of the
integrals. The corresponding initial values (\ref{in-1})--(\ref{in0}) can be
found in Ref.~\cite{Suslov} or elsewhere.\medskip 

A different representation of these integrals in terms of Chebyshev
polynomials is also available \cite{Suslov}:%
\begin{eqnarray}
&&2a\mu \left( 2a\beta \right) ^{p}\ A_{p}  \label{meAha} \\
&&\quad =a\left( \mu +a\kappa \right) \ h_{p}^{\left( 0,\ 0\right) }\left(
n-1,-2\nu \right)  \notag \\
&&\quad +2\varepsilon \left( \mu ^{2}-a^{2}\kappa ^{2}\right) \
h_{p-1}^{\left( 1,\ 1\right) }\left( n-1,-1-2\nu \right)  \notag \\
&&\quad +a\left( \mu -a\kappa \right) \ h_{p}^{\left( 0,\ 0\right) }\left(
n,-2\nu \right) ,  \notag
\end{eqnarray}%
\begin{eqnarray}
&&2a\mu \left( 2a\beta \right) ^{p}\ B_{p}  \label{meBha} \\
&&\quad =a\varepsilon \left( \mu +a\kappa \right) \ h_{p}^{\left( 0,\
0\right) }\left( n-1,-2\nu \right)  \notag \\
&&\quad +2\left( \mu ^{2}-a^{2}\kappa ^{2}\right) \ h_{p-1}^{\left( 1,\
1\right) }\left( n-1,-1-2\nu \right)  \notag \\
&&\quad +a\varepsilon \left( \mu -a\kappa \right) \ h_{p}^{\left( 0,\
0\right) }\left( n,-2\nu \right) ,  \notag
\end{eqnarray}%
\begin{eqnarray}
&&4\mu \left( 2a\beta \right) ^{p}\ C_{p}  \label{meCha} \\
&&\quad =a\varepsilon \left( \mu +a\kappa \right) \ h_{p}^{\left( 0,\
0\right) }\left( n-1,-2\nu \right)  \notag \\
&&\quad -a\varepsilon \left( \mu -a\kappa \right) \ h_{p}^{\left( 0,\
0\right) }\left( n,-2\nu \right) ,  \notag
\end{eqnarray}%
when $p\geq 1.$ A similar matrix manipulation results in another recurrence
relation:%
\begin{eqnarray}
\left( 2a\beta \right) ^{2}\ A_{p+1} &=&4\beta \varepsilon \mu \frac{\left(
a^{2}\left( p+1\right) ^{2}-1\right) \left( 2p+1\right) }{a^{2}p\left(
p+1\right) \left( p+2\right) }\ A_{p}  \label{meACheb} \\
&&+\frac{4\beta \varepsilon ^{2}\mu \left( 2p+1\right) }{a^{2}p\left(
p+1\right) \left( p+2\right) }\ B_{p}-4\beta \frac{2p+1}{p+1}\ C_{p}  \notag
\\
&&-\frac{\left( a^{2}\left( p+1\right) ^{2}-1\right) \left( 4\nu
^{2}-p^{2}\right) }{a^{2}\left( p+1\right) \left( p+2\right) }\ A_{p-1}-%
\frac{\varepsilon \left( 4\nu ^{2}-p^{2}\right) }{a^{2}\left( p+1\right)
\left( p+2\right) }\ B_{p-1},  \notag
\end{eqnarray}%
\begin{eqnarray}
\left( 2a\beta \right) ^{2}\ B_{p+1} &=&-\frac{4\beta \varepsilon ^{2}\mu
\left( 2p+1\right) }{a^{2}p\left( p+1\right) \left( p+2\right) }A_{p}
\label{meBCheb} \\
&&+4\beta \varepsilon \mu \frac{\left( a^{2}\left( p+1\right)
^{2}+\varepsilon ^{2}\right) \left( 2p+1\right) }{a^{2}p\left( p+1\right)
\left( p+2\right) }\ B_{p}-4\beta \varepsilon \frac{2p+1}{p+1}\ C_{p}  \notag
\\
&&+\frac{\varepsilon \left( 4\nu ^{2}-p^{2}\right) }{a^{2}\left( p+1\right)
\left( p+2\right) }\ A_{p-1}-\frac{\left( 4\nu ^{2}-p^{2}\right) \left(
a^{2}\left( p+1\right) ^{2}+\varepsilon ^{2}\right) }{a^{2}\left( p+1\right)
\left( p+2\right) }\ B_{p-1},  \notag
\end{eqnarray}%
\begin{eqnarray}
\left( 2a\beta \right) ^{2}\ C_{p+1} &=&-\beta \frac{2p+1}{p+1}A_{p}+\beta
\varepsilon \frac{2p+1}{p+1}\ B_{p}  \label{meCCheb} \\
&&+4\beta \varepsilon \mu \frac{2p+1}{p+1}\ C_{p}-\left( 4\nu
^{2}-p^{2}\right) \frac{p}{p+1}\ C_{p-1}.  \notag
\end{eqnarray}%
These recurrence relations are summarized for the benefit of the reader in
matrix form in appendix~A. One should take special values (\ref{in0})--(\ref%
{in1}) as the initial data. A direct derivation of the relativistic
three-term vector recurrence relations (\ref{4termA})--(\ref{4termC}) and (%
\ref{meACheb})--(\ref{meCCheb}) on the basis of a hypervirial theorem needs
to be found.

\section{Independent Integrals}

The integrals (\ref{meA})--(\ref{meC}) are linearly dependent: 
\begin{equation}
\left( 2\kappa +\varepsilon \left( p+1\right) \right) A_{p}-\left(
2\varepsilon \kappa +p+1\right) B_{p}=4\mu C_{p}  \label{indint1}
\end{equation}%
(see, for example, \cite{Adkins}, \cite{ShabVest}, \cite{Shab91}, and \cite%
{Suslov} for more details). Thus, eliminating $C_{p},$ say from (\ref%
{meACheb})--(\ref{meBCheb}), we obtain the following three-term vector
recurrence relation between the integrals $A_{p}$ and $B_{p}$ only: 
\begin{eqnarray}
&&\left( 2a\beta \right) ^{2}\ A_{p+1}  \label{2DA} \\
&&\quad =\left( 2p+1\right) \beta \frac{4\varepsilon \mu ^{2}\left(
a^{2}\left( p+1\right) ^{2}-1\right) -a^{2}p\left( p+2\right) \left( 2\kappa
+\varepsilon \left( p+1\right) \right) }{a^{2}\mu p\left( p+1\right) \left(
p+2\right) }\ A_{p}  \notag \\
&&\qquad +\left( 2p+1\right) \beta \frac{4\varepsilon ^{2}\mu
^{2}+a^{2}p\left( p+2\right) \left( 2\varepsilon \kappa +p+1\right) }{%
a^{2}\mu p\left( p+1\right) \left( p+2\right) }\ B_{p}  \notag \\
&&\qquad -\frac{\left( a^{2}\left( p+1\right) ^{2}-1\right) \left( 4\nu
^{2}-p^{2}\right) }{a^{2}\left( p+1\right) \left( p+2\right) }\ A_{p-1}-%
\frac{\varepsilon \left( 4\nu ^{2}-p^{2}\right) }{a^{2}\left( p+1\right)
\left( p+2\right) }\ B_{p-1}  \notag
\end{eqnarray}%
and%
\begin{eqnarray}
&&\left( 2a\beta \right) ^{2}\ B_{p+1}  \label{2DB} \\
&&\quad =-\left( 2p+1\right) \beta \varepsilon \frac{4\varepsilon \mu
^{2}+a^{2}p\left( p+2\right) \left( 2\kappa +\varepsilon \left( p+1\right)
\right) }{a^{2}\mu p\left( p+1\right) \left( p+2\right) }A_{p}  \notag \\
&&\qquad +\left( 2p+1\right) \beta \varepsilon \frac{4\mu ^{2}\left(
a^{2}\left( p+1\right) ^{2}+\varepsilon ^{2}\right) +a^{2}p\left( p+2\right)
\left( 2\varepsilon \kappa +p+1\right) }{a^{2}\mu p\left( p+1\right) \left(
p+2\right) }\ B_{p}  \notag \\
&&\qquad +\frac{\varepsilon \left( 4\nu ^{2}-p^{2}\right) }{a^{2}\left(
p+1\right) \left( p+2\right) }\ A_{p-1}-\frac{\left( 4\nu ^{2}-p^{2}\right)
\left( a^{2}\left( p+1\right) ^{2}+\varepsilon ^{2}\right) }{a^{2}\left(
p+1\right) \left( p+2\right) }\ B_{p-1}.  \notag
\end{eqnarray}%
In a similar fashion, from (\ref{4termA})--(\ref{4termB}):%
\begin{eqnarray}
&&\left( 2a\beta \right) ^{2}\ A_{p+1}=4\beta \varepsilon \mu \frac{2p+3}{p+2%
}\ A_{p}  \label{2DAA} \\
&&\quad +p\frac{2\varepsilon \kappa -\left( p+2\right) \left( 4\nu
^{2}-\left( p+1\right) ^{2}\right) }{\left( p+1\right) \left( p+2\right) }\
A_{p-1}  \notag \\
&&\qquad -2\kappa \frac{2\varepsilon \kappa -1+\left( p+1\right) \left(
2p+3\right) }{\left( p+1\right) \left( p+2\right) }\ B_{p-1}  \notag
\end{eqnarray}%
and%
\begin{eqnarray}
&&\left( 2a\beta \right) ^{2}\ B_{p+1}=4\beta \varepsilon \mu \frac{2p+3}{p+2%
}\ B_{p}  \label{2DBB} \\
&&\quad +2\varepsilon p\frac{p+1}{p+2}\ A_{p-1}-\left( 4\left( \varepsilon
\kappa +\nu ^{2}\right) -p^{2}\right) \frac{p+1}{p+2}\ B_{p-1}.  \notag
\end{eqnarray}%
Matrix forms of these identities and the corresponding initial values are
given in appendix~A.

\section{Two-Term Recurrence Relations}

The three-term recurrence relations for the relativistic radial integrals,
examples of which have been found in the previous sections, are obviously
not unique. Moreover, if%
\begin{eqnarray}
\mathbf{A}_{p+1} &=&D_{p}^{\left( 1\right) }\mathbf{A}_{p}+E_{p}^{\left(
1\right) }\mathbf{A}_{p-1},  \label{3term2} \\
\mathbf{A}_{p+1} &=&D_{p}^{\left( 2\right) }\mathbf{A}_{p}+E_{p}^{\left(
2\right) }\mathbf{A}_{p-1},  \notag
\end{eqnarray}%
then%
\begin{eqnarray}
\left( \alpha _{p}+\beta _{p}\right) \mathbf{A}_{p+1} &=&\left( \alpha
_{p}D_{p}^{\left( 1\right) }+\beta _{p}D_{p}^{\left( 2\right) }\right) 
\mathbf{A}_{p}  \label{3term1} \\
&&+\left( \alpha _{p}E_{p}^{\left( 1\right) }+\beta _{p}E_{p}^{\left(
2\right) }\right) \mathbf{A}_{p-1}  \notag
\end{eqnarray}%
for two arbitrary sequences of scalars $\alpha _{p}$ and $\beta _{p}.$ One
special case is of a particular interest.\medskip

Subtracting (\ref{2DA})--(\ref{2DB}) from (\ref{2DAA})--(\ref{2DBB}) one
gets the following matrix equation%
\begin{equation}
P_{p}\ \left( 
\begin{array}{c}
A_{p}\medskip \\ 
B_{p}%
\end{array}%
\right) =Q_{p}\ \left( 
\begin{array}{c}
A_{p-1}\medskip \\ 
B_{p-1}%
\end{array}%
\right)  \label{PQ}
\end{equation}%
with%
\begin{eqnarray}
\det P_{p} &=&-8\beta ^{2}\varepsilon \frac{\kappa a^{2}\left( p+2\right)
\left( 2p+1\right) +2\varepsilon \mu ^{2}}{p\left( p+1\right) \left(
p+2\right) ^{2}},  \label{det1} \\
\det Q_{p} &=&-2\varepsilon \left( 4\nu ^{2}-p^{2}\right) \frac{\kappa
a^{2}\left( p+2\right) \left( 2p+1\right) +2\varepsilon \mu ^{2}}{%
a^{2}\left( p+2\right) ^{2}\left( p+1\right) ^{2}}  \label{det2}
\end{eqnarray}%
by a computer algebra system. We have omitted the explicit forms of the $P$
and $Q$ matrices and their inverses. It should be easily done by the reader.

The two-term recurrence solutions of the form%
\begin{equation}
\left( 
\begin{array}{c}
A_{p}\medskip \\ 
B_{p}%
\end{array}%
\right) =S_{p}\left( 
\begin{array}{c}
A_{p-1}\medskip \\ 
B_{p-1}%
\end{array}%
\right) ,\qquad \left( 
\begin{array}{c}
A_{p-1}\medskip \\ 
B_{p-1}%
\end{array}%
\right) =S_{p}^{-1}\left( 
\begin{array}{c}
A_{p}\medskip \\ 
B_{p}%
\end{array}%
\right)  \label{MatSol}
\end{equation}%
where found by Shabaev \cite{ShabVest}, \cite{Shab91} by a different method.
In our notations,%
\begin{eqnarray}
A_{p+1} &=&-\left( p+1\right) \frac{4\nu ^{2}\varepsilon +2\kappa \left(
p+2\right) +\varepsilon \left( p+1\right) \left( 2\kappa \varepsilon
+p+2\right) }{4\left( 1-\varepsilon ^{2}\right) \left( p+2\right) \beta \mu }%
\ A_{p}  \label{rra} \\
&&+\frac{4\mu ^{2}\left( p+2\right) +\left( p+1\right) \left( 2\kappa
\varepsilon +p+1\right) \left( 2\kappa \varepsilon +p+2\right) }{4\left(
1-\varepsilon ^{2}\right) \left( p+2\right) \beta \mu }\ B_{p},  \notag
\end{eqnarray}%
\begin{eqnarray}
B_{p+1} &=&-\left( p+1\right) \frac{4\nu ^{2}+2\kappa \varepsilon \left(
2p+3\right) +\varepsilon ^{2}\left( p+1\right) \left( p+2\right) }{4\left(
1-\varepsilon ^{2}\right) \left( p+2\right) \beta \mu }\ A_{p}  \label{rrb}
\\
&&+\frac{4\mu ^{2}\varepsilon \left( p+2\right) +\left( p+1\right) \left(
2\kappa \varepsilon +p+1\right) \left( 2\kappa +\varepsilon \left(
p+2\right) \right) }{4\left( 1-\varepsilon ^{2}\right) \left( p+2\right)
\beta \mu }\ B_{p}  \notag
\end{eqnarray}%
and%
\begin{eqnarray}
A_{p-1} &=&\beta \frac{4\mu ^{2}\varepsilon \left( p+1\right) +p\left(
2\kappa \varepsilon +p\right) \left( 2\kappa +\varepsilon \left( p+1\right)
\right) }{\mu \left( 4\nu ^{2}-p^{2}\right) p}\ A_{p}  \label{rrab} \\
&&-\beta \frac{4\mu ^{2}\left( p+1\right) +p\left( 2\kappa \varepsilon
+p\right) \left( 2\kappa \varepsilon +p+1\right) }{\mu \left( 4\nu
^{2}-p^{2}\right) p}\ B_{p},  \notag
\end{eqnarray}%
\begin{eqnarray}
B_{p-1} &=&\beta \frac{4\nu ^{2}+2\kappa \varepsilon \left( 2p+1\right)
+\varepsilon ^{2}p\left( p+1\right) }{\mu \left( 4\nu ^{2}-p^{2}\right) }\
A_{p}  \label{rrba} \\
&&-\beta \frac{4\nu ^{2}\varepsilon +2\kappa \left( p+1\right) +\varepsilon
p\left( 2\kappa \varepsilon +p+1\right) }{\mu \left( 4\nu ^{2}-p^{2}\right) }%
\ B_{p},  \notag
\end{eqnarray}%
\medskip respectively.

We have shown that these solutions can be derived from the three-term vector
recurrence relations found in this paper. As a by-product, the factorization
of Shabaev's matrices, namely,%
\begin{equation}
S_{p}=P_{p}^{-1}\ Q_{p},\qquad S_{p}^{-1}=Q_{p}^{-1}\ P_{p},  \label{factor}
\end{equation}%
is given. This can be directly verified with the help of a computer algebra
system. Then%
\begin{equation}
\det S_{p}=\frac{\det Q_{p}}{\det P_{p}}=\frac{\left( 4\nu ^{2}-p^{2}\right)
p}{4\left( a\beta \right) ^{2}\left( p+1\right) }.  \label{dets}
\end{equation}%
Further details are left to the reader.\medskip

\noindent \textbf{Acknowledgments.\/} I thank Carlos Castillo-Ch\'{a}vez for
support and encouragement and David Murillo for help. I am grateful to
Vladimir M.~Shabaev for a copy of \cite{ShabVest}. The referees' suggestions
are also very much appreciated.

\appendix

\section{Matrix Form of Three-Term Recurrence Relations}

The matrix structures of our recurrence relations (\ref{4termA})--(\ref%
{4termC}) and (\ref{meACheb})--(\ref{meCCheb}) are given by%
\begin{eqnarray}
&&\left( 2a\beta \right) ^{2}\left( 
\begin{array}{c}
A_{p+1}\smallskip  \\ 
B_{p+1}\smallskip  \\ 
C_{p+1}%
\end{array}%
\right) =4\beta \varepsilon \mu \frac{2p+3}{p+2}\left( 
\begin{array}{c}
A_{p}\smallskip  \\ 
B_{p}\smallskip  \\ 
C_{p}%
\end{array}%
\right)   \label{mat1} \\
&&\qquad -\left( 
\begin{array}{ccc}
\frac{p\left( p+2\right) \left( 4\nu ^{2}-\left( p+1\right) ^{2}\right)
+4\kappa ^{2}}{\left( p+1\right) \left( p+2\right) } & 4\kappa \frac{p+1}{p+2%
} & -\frac{8\kappa \mu }{\left( p+1\right) \left( p+2\right) }\smallskip  \\ 
4\kappa \frac{p+1}{p+2} & \left( 4\nu ^{2}-p\left( p+2\right) \right) \frac{%
p+1}{p+2} & -8\mu \frac{p+1}{p+2}\smallskip  \\ 
\frac{2\kappa \mu }{\left( p+1\right) \left( p+2\right) } & 2\mu \frac{p+1}{%
p+2} & \frac{p\left( p+2\right) \left( 4\nu ^{2}-\left( p+1\right)
^{2}\right) -4\mu ^{2}}{\left( p+1\right) \left( p+2\right) }%
\end{array}%
\right) \left( 
\begin{array}{c}
A_{p-1}\smallskip  \\ 
B_{p-1}\smallskip  \\ 
C_{p-1}%
\end{array}%
\right)   \notag
\end{eqnarray}%
and%
\begin{eqnarray}
&&\left( 2a\beta \right) ^{2}\left( 
\begin{array}{c}
A_{p+1}\smallskip  \\ 
B_{p+1}\smallskip  \\ 
C_{p+1}%
\end{array}%
\right)   \label{mat2} \\
&&\quad =\beta \frac{2p+1}{p+1}\left( 
\begin{array}{ccc}
\frac{4\varepsilon \mu \left( a^{2}\left( p+1\right) ^{2}-1\right) }{%
a^{2}p\left( p+2\right) } & \frac{4\varepsilon ^{2}\mu }{a^{2}p\left(
p+2\right) } & -4\smallskip  \\ 
-\frac{4\varepsilon ^{2}\mu }{a^{2}p\left( p+2\right) } & \frac{4\varepsilon
\mu \left( a^{2}\left( p+1\right) ^{2}+\varepsilon ^{2}\right) }{%
a^{2}p\left( p+2\right) } & -4\varepsilon \smallskip  \\ 
-1 & \varepsilon  & 4\varepsilon \mu 
\end{array}%
\right) \left( 
\begin{array}{c}
A_{p}\smallskip  \\ 
B_{p}\smallskip  \\ 
C_{p}%
\end{array}%
\right)   \notag \\
&&\qquad -\frac{\left( 4\nu ^{2}-p^{2}\right) }{p+1}\left( 
\begin{array}{ccc}
\frac{\left( a^{2}\left( p+1\right) ^{2}-1\right) }{a^{2}\left( p+2\right) }
& \frac{\varepsilon }{a^{2}\left( p+2\right) } & 0\smallskip  \\ 
-\frac{\varepsilon }{a^{2}\left( p+2\right) } & \frac{\left( a^{2}\left(
p+1\right) ^{2}+\varepsilon ^{2}\right) }{a^{2}\left( p+2\right) } & 
0\smallskip  \\ 
0 & 0 & p%
\end{array}%
\right) \left( 
\begin{array}{c}
A_{p-1}\smallskip  \\ 
B_{p-1}\smallskip  \\ 
C_{p-1}%
\end{array}%
\right) ,  \notag
\end{eqnarray}%
respectively. The initial vectors are%
\begin{equation}
\left( 
\begin{array}{c}
A_{-1}\smallskip  \\ 
B_{-1}\smallskip  \\ 
C_{-1}%
\end{array}%
\right) =\left( 
\begin{array}{c}
\frac{\beta }{\mu \nu }\left( 1-\varepsilon ^{2}\right) \left( \varepsilon
\nu +\mu \sqrt{1-\varepsilon ^{2}}\right) \smallskip  \\ 
\frac{\beta a^{2}}{\mu }\smallskip  \\ 
\frac{\kappa }{2\mu \nu }a^{3}\beta 
\end{array}%
\right) ,  \label{in-1}
\end{equation}%
\begin{equation}
\left( 
\begin{array}{c}
A_{0}\smallskip  \\ 
B_{0}\smallskip  \\ 
C_{0}%
\end{array}%
\right) =\left( 
\begin{array}{c}
1\smallskip  \\ 
\varepsilon \smallskip  \\ 
\frac{\kappa }{2\mu }\left( 1-\varepsilon ^{2}\right) 
\end{array}%
\right) ,  \label{in0}
\end{equation}%
\begin{equation}
\left( 
\begin{array}{c}
A_{1}\smallskip  \\ 
B_{1}\smallskip  \\ 
C_{1}%
\end{array}%
\right) =\left( 
\begin{array}{c}
\frac{3\varepsilon \mu ^{2}-\kappa \left( 1-\varepsilon ^{2}\right) \left(
1+\varepsilon \kappa \right) }{2\beta \mu \left( 1-\varepsilon ^{2}\right) }%
\smallskip  \\ 
\frac{3\varepsilon ^{2}\mu ^{2}-\left( 1-\varepsilon ^{2}\right) \left(
\varepsilon \kappa +\nu ^{2}\right) }{2\beta \mu \left( 1-\varepsilon
^{2}\right) }\smallskip  \\ 
\frac{2\varepsilon \kappa -1}{4\beta }%
\end{array}%
\right) .  \label{in1}
\end{equation}%
One should use the initial data (\ref{in-1})--(\ref{in0}) in the case of the
recurrence relation (\ref{mat1}) and (\ref{in0})--(\ref{in1}) for (\ref{mat2}%
).\medskip 

Our relations (\ref{2DA})--(\ref{2DB}) and (\ref{2DAA})--(\ref{2DBB}) take
the following matrix forms:%
\begin{eqnarray}
&&\left( 2a\beta \right) ^{2}\left( 
\begin{array}{c}
A_{p+1}\medskip \\ 
B_{p+1}%
\end{array}%
\right)  \label{mat3} \\
&&\quad =\beta \frac{2p+1}{p+1}\left( 
\begin{array}{cc}
\frac{4\varepsilon \mu ^{2}\left( a^{2}\left( p+1\right) ^{2}-1\right)
-a^{2}p\left( p+2\right) \left( 2\kappa +\varepsilon \left( p+1\right)
\right) }{a^{2}\mu p\left( p+2\right) } & \frac{4\varepsilon ^{2}\mu
^{2}+a^{2}p\left( p+2\right) \left( 2\varepsilon \kappa +p+1\right) }{%
a^{2}\mu p\left( p+2\right) }\medskip \\ 
-\varepsilon \frac{4\varepsilon \mu ^{2}+a^{2}p\left( p+2\right) \left(
2\kappa +\varepsilon \left( p+1\right) \right) }{a^{2}\mu p\left( p+2\right) 
} & \varepsilon \frac{4\mu ^{2}\left( a^{2}\left( p+1\right)
^{2}+\varepsilon ^{2}\right) +a^{2}p\left( p+2\right) \left( 2\varepsilon
\kappa +p+1\right) }{a^{2}\mu p\left( p+2\right) }%
\end{array}%
\right)  \notag \\
&&\qquad \times \left( 
\begin{array}{c}
A_{p}\medskip \\ 
B_{p}%
\end{array}%
\right) -\frac{\left( 4\nu ^{2}-p^{2}\right) }{a^{2}\left( p+1\right) \left(
p+2\right) }\left( 
\begin{array}{cc}
a^{2}\left( p+1\right) ^{2}-1 & \medskip \varepsilon \\ 
-\varepsilon & a^{2}\left( p+1\right) ^{2}+\varepsilon ^{2}%
\end{array}%
\right) \left( 
\begin{array}{c}
A_{p-1}\medskip \\ 
B_{p-1}%
\end{array}%
\right)  \notag
\end{eqnarray}%
and%
\begin{eqnarray}
&&\left( 2a\beta \right) ^{2}\left( 
\begin{array}{c}
A_{p+1}\medskip \\ 
B_{p+1}%
\end{array}%
\right) =4\beta \varepsilon \mu \frac{2p+3}{p+2}\left( 
\begin{array}{c}
A_{p}\medskip \\ 
B_{p}%
\end{array}%
\right)  \label{mat4} \\
&&\quad -\left( 
\begin{array}{cc}
-p\frac{2\varepsilon \kappa -\left( p+2\right) \left( 4\nu ^{2}-\left(
p+1\right) ^{2}\right) }{\left( p+1\right) \left( p+2\right) } & 2\kappa 
\frac{2\varepsilon \kappa -1+\left( p+1\right) \left( 2p+3\right) }{\left(
p+1\right) \left( p+2\right) }\medskip \\ 
-2\varepsilon p\frac{p+1}{p+2} & \left( 4\left( \varepsilon \kappa +\nu
^{2}\right) -p^{2}\right) \frac{p+1}{p+2}%
\end{array}%
\right) \left( 
\begin{array}{c}
A_{p-1}\medskip \\ 
B_{p-1}%
\end{array}%
\right) ,  \notag
\end{eqnarray}%
respectively.

\end{document}